\documentclass[10pt,b5paper,onecolumn,twoside]{strat-contribution}
\usepackage{defsContribution}
\usepackage{amsmath,amssymb,enumitem,float,relsize}
\usepackage{datetime}
\usepackage[pdftex]{graphicx}

\setTitle{A modal logic translation of the AGM \linebreak
axioms for belief revision}

\setShortTitle{Modal AGM}

\setAuthor{Giacomo Bonanno
}

\setShortAuthor{G Bonanno}

\setInstitute{Department of Economics, University of California, Davis, USA}

\setEmail{gfbonanno@ucdavis.edu}

\begin{document}

\maketitle

\begin{abstract}
Building on the analysis of \cite{Bon25a} we introduce a simple modal logic containing three modal operators: a unimodal belief operator $B$, a bimodal conditional operator $>$ and the unimodal global operator $\square$. For each AGM axiom for belief revision, we provide a corresponding modal axiom. The correspondence is as follows: each AGM axiom is characterized by a property of the Kripke-Lewis frames considered in \cite{Bon25a} and, in turn, that property characterizes the proposed modal axiom.
\end{abstract}

\begin{center}
\newdateformat{specialdate}{\small{\monthname\,\,\THEDAY,\,\,\THEYEAR}}
\date{\specialdate\today}
\end{center}
\section{Introduction}
In \cite{Bon25a} a new semantics for both belief update and belief revision was provided in terms of frames consisting of a set of states, a Kripke belief relation and a Lewis selection function. In this paper we focus on AGM belief revision and make use of that semantics to establish a correspondence between each AGM axiom and a modal axiom in a logic that contains three modal operators: a unimodal belief operator $B$, a bimodal conditional operator $>$ and the unimodal global operator $\square$. Adding a valuation to a frame yields a model. Given a model and a state $s$, we identify the initial belief set $K$ with the set of formulas that are believed at state $s$ (that is, $K=\{\phi:s\models B\phi\}$) and we identify the revised belief set $K\ast\phi$ (prompted by the input represented by formula $\phi$) as the set of formulas that are the consequent of conditionals that (1) are believed at state $s$ and (2) have $\phi$ as antecedent (that is, $K\ast\phi=\{\psi:s\models B(\phi>\psi)\}$). The next section briefly reviews the AGM axioms for belief revision and the approach put forward in  \cite{Bon25a}, while Section 3 introduces the modal logic and provides axioms that correspond to the semantic properties that characterize the AGM axioms. Thus there is a precise sense in which each proposed modal axiom corresponds to the respective AGM axiom.
\section{AGM axioms and their semantic characterization}
\label{Sec2}
We consider a propositional logic based on a countable set \texttt{At} of atomic formulas. We denote by $\Phi_0$ the set of Boolean formulas constructed from  \texttt{At} as follows: $\texttt{At}\subset \Phi_0$ and if $\phi,\psi\in\Phi_0$ then $\neg\phi$ and $\phi\vee\psi$ belong to $\Phi_0$. Define  $\phi\rightarrow\psi$, $\phi\wedge\psi$, and $\phi\leftrightarrow\psi$ in terms of $\neg$ and $\vee$ in the usual way.
\par
Given a subset $K$ of $\Phi_0 $, its deductive closure $Cn(K)\subseteq\Phi_0$ is defined as follows: $\psi \in Cn(K)$ if and only if there exist $\phi _1,...,\phi _n\in K$ \ (with $n\geq 0$) such that $(\phi _1\wedge ...\wedge \phi _n)\rightarrow \psi $ is a tautology. A set $K\subseteq \Phi_0 $ is \textit{consistent} if $ Cn(K)\neq \Phi_0 $; it is \textit{deductively closed} if $K=Cn(K)$. Given a set $K\subseteq \Phi_0$ and a formula $\phi\in\Phi_0$, the \emph{expansion} of $K$ by $\phi$, denoted by $K+\phi$, is defined as follows: $K+\phi=Cn\left(K\cup\{\phi\}\right)$.
\par
Let $K\subseteq\Phi_0$ be a \emph{consistent and deductively closed} set representing the agent's initial beliefs and let $\Psi \subseteq \Phi_0 $ be a set of formulas representing possible inputs for belief change. A \emph{belief change function} based on $\Psi$ and $K$ is a function $\circ:\Psi \rightarrow 2^{\Phi_0 }$ ($2^{\Phi_0}$ denotes the set of subsets of $\Phi_0 $) that associates with every formula $\phi \in \Psi $ a set $ K\circ\phi \subseteq \Phi_0 $, interpreted as the change in $K$ prompted by the input $\phi$. We follow the common practice of writing $K\circ\phi$ instead of $\circ(\phi)$ which has the advantage of making it clear that the belief change function refers to a given, \emph{fixed}, $K$. If $\Psi \neq \Phi_0 $ then $\circ$ is called a \emph{partial} belief change function, while if $\Psi =\Phi_0 $ then $\circ$ is called a \emph{full-domain} belief change function.
\begin{definition}
\label{brf extension}
Let $\circ:\Psi \rightarrow 2^{\Phi_0}$ be a partial belief change function (thus $\Psi\ne\Phi_0$) and $\circ ':\Phi_0 \rightarrow 2^{\Phi_0}$ a full-domain belief change function. We say that $\circ '$ is an \emph{extension} of $\circ$ if $\circ '$ coincides with $\circ $ on the domain of $\circ $, that is, if, for every $\phi \in \Psi$, $ K\circ '\phi= K\circ\phi $.
\end{definition}
\subsection{AGM axioms}
We consider the notion of belief revision proposed Alchourr\'{o}n, G\"{a}rdenfors and Makinson in \cite{AGM85}.
\begin{definition}
\label{DEF:revision}
A \emph{belief revision function}, based on the  consistent and deductively closed set $K\subset \Phi_0$ (representing the initial beliefs), is a full domain belief change function $\ast:\Phi_0\to 2^{\Phi_0}$ that satisfies the following axioms: $\forall \phi,\psi\in\Phi_0$,
\begin{enumerate}
\item[]($K\ast 1$)\quad $K\ast\phi=Cn(K\ast\phi)$.
\item[]($K\ast 2$)\quad $\phi \in K\ast\phi$.
\item[]($K\ast 3$)\quad $K\ast\phi\subseteq K+ \phi$.
\item[]($K\ast 4$)\quad if $\neg \phi \notin K$, then $K\subseteq K\ast\phi$.
\item[]($K\ast 5$)\quad $\begin{array}{ll}(K\ast 5a)&\text{If }\neg\phi \text{ is a tautology, then } K\ast\phi=\Phi_0.\\[3pt]
     (K\ast 5b)&\text{If }\neg\phi \text{ is not a tautology, then } K\ast\phi\ne\Phi_0.\end{array}$
\item[]($K\ast 6$)\quad if $\phi \leftrightarrow \psi $ is a tautology then $K\ast\phi=K\ast\psi$.
\item[]($K\ast 7$)\quad $K\ast(\phi \wedge \psi)\subseteq (K\ast\phi)+\psi$.
\item[]($K\ast 8$)\quad if $\neg \psi \notin K\ast\phi$, then $(K\ast\phi)+\psi\subseteq K\ast(\phi \wedge \psi).$
\end{enumerate}
\end{definition}
$K\ast\phi$ is interpreted as the revised belief set in response to receiving the input represented by formula $\phi$. For a discussion of the above axioms, known as the AGM axioms, see, for example \cite{FerHans18,Gae88}.
\subsection{Kripke-Lewis semantics}
\label{Sec2.2}
\begin{definition}
\label{DEF:frame}
A \emph{Kripke-Lewis frame} is a triple $\left\langle {S,\mathcal B,f} \right\rangle$ where
\begin{enumerate}
\item $S$ is a set of \emph{states}; subsets of $S$ are called  \emph{events}.
\item $\mathcal B \subseteq S \times S$ is a binary \emph{belief relation} on $S$ which is serial: $\forall s\in S, \exists s'\in S$, such that $s\mathcal B s'$ ($s\mathcal B s'$ is an alternative notation for $(s,s')\in\mathcal B$). We denote by $\mathcal B(s)$ the set of states that are reachable from $s$ by $\mathcal B$: $\mathcal B(s)=\{s'\in S: s\mathcal B s'\}$. $\mathcal B(s)$ is interpreted as the set of states that initially the agent considers doxastically possible at state $s$.
\item $f:S\times (2^S\setminus\{\varnothing\}) \rightarrow 2^S$ is a \emph{selection function} that associates with every state-event pair $(s,E)$ (with $E\ne\varnothing$) a set of states $f(s,E)\subseteq S$.\footnote{Note that $f(s,E)$ is defined only if $E\ne\varnothing$. The reason for this will become clear later. In \cite{Bon25a} the selection function was assumed to satisfy the following standard properties:
    \begin{enumerate}
    \item[] (3.1)\quad $ f(s,E)\ne\varnothing$ (Consistency).
    \item[] (3.2)\quad $ f(s,E)\subseteq E$ (Success).
    \item[] (3.3)\quad if $s\in E$ then $s\in f(s,E)$ (Weak Centering).
    \end{enumerate}
    Here we do not require any properties to start with, because we want to highlight the role of each frame property in the characterization of the AGM axioms.}
    \par\noindent
    $ f(s,E)$ is interpreted as the set of states that are closest (or most similar) to $s$, conditional on event $E$.
    \end{enumerate}
\end{definition}
Adding a valuation to a frame yields a model. Thus a \emph{model} is a tuple $\left\langle {S,\mathcal B,f,V} \right\rangle$ where $\left\langle {S,\mathcal B,f} \right\rangle$ is a frame and $V:\texttt{At}\rightarrow 2^S$ is a valuation that assigns to every atomic formula $p\in\texttt{At}$ the set of states where $p$ is true.
\begin{definition}
\label{Truth0}
Given a model $M=\left\langle {S,\mathcal B,f,V} \right\rangle$ define truth of a Boolean formula $\phi\in\Phi_0$ at a state $s\in S$ in model $M$, denoted by $s\models_M\phi$, as follows:
\begin{enumerate}
\item if $p\in\texttt{At}$ then $s\models_M p$ if and only if $s\in V(p)$,
\item $s\models_M\neg\phi$ if and only if $s\not\models_M\phi$,
\item $s\models_M(\phi\vee\psi)$ if and only if $s\models_M\phi$ or $s\models_M\psi$ (or both).
\end{enumerate}
\end{definition}
\noindent
We denote by $\Vert\phi\Vert_M$ the truth set of formula $\phi$ in model $M$: $$\Vert\phi\Vert_M=\{s\in S: s\models_M\phi\}.$$
\par
Given a model $M=\left\langle {S,\mathcal B,f,V} \right\rangle$ and a state $s\in S$,  let $K_{s,M}=\{\phi \in \Phi _{0}:\mathcal B(s) \subseteq \Vert \phi \Vert_{M} \}$; thus a Boolean formula $\phi$ belongs to $K_{s,M}$ if and only if at state $s$ the agent believes $\phi$ (in the sense that $\phi$ is true at every state that the agent considers doxastically possible at state $s$). We identify $K_{s,M}$ with the agent's \emph{initial beliefs at state} $s$. It is shown in \cite{Bon25a} that the set $K_{s,M}\subseteq\Phi_0$ so defined is deductively closed and consistent (recall the assumption that the belief relation $\mathcal B$ is serial).
\par
 Next, given a model $M=\left\langle {S,\mathcal B,f,V} \right\rangle$ and a state $s\in S$, let $\Psi_M=\{\phi\in\Phi_0:\Vert\phi\Vert_M\ne\varnothing\}$\footnote{%%%% FOOTNOTE %%%%
Since, in any given model there are formulas $\phi$ such that $\Vert\phi\Vert_M=\varnothing$ (at the very least all the contradictions), $\Psi_M$ is a proper subset of $\Phi_0$.
  }\ %END FOOTNOTE%%%
and define the following \emph{partial} belief change function
 $\circ:\Psi_M\to 2^{\Phi_0}$ based on $K_{s,M}$:
\begin{equation}
\tag{RI}
\label{RI}
\begin{array}{*{20}{l}}
 \psi\in K_{s,M}\circ \phi\,\,\text{ if and only if, }\,\, \forall s'\in\mathcal B(s), \, f\left(s',\Vert\phi\Vert_M\right)\subseteq\Vert\psi\Vert_M.
\end{array}
\end{equation}
Given the customary interpretation of selection functions in terms of conditionals, (RI) can be interpreted as stating that $\psi\in K_{s,M}\circ\phi$ if and only if at state $s$ the agent believes that "if $\phi$ is (were) the case then $\psi$ is (would be) the case". This interpretation will be made explicit in the modal logic considered in Section 3.
In what follows, when stating an axiom for a belief change function, we implicitly assume that it applies to every formula \emph{in its domain}. For example, the axiom $\phi\in K\circ\phi$ asserts that, for all $\phi$ in the domain of $\circ$, \,$\phi\in K\circ\phi$.
\begin{definition}
\label{DEF:validity}
An axiom for belief change functions is \emph{valid on a frame $F$} if, for every model based on that frame and for every state $s$ in that model, the partial belief change function defined by \eqref{RI} satisfies the axiom. An axiom is \emph{valid on a set of frames $\mathcal F$} if it is valid on every frame $F\in\mathcal F$.
\end{definition}
\subsection{Frame correspondence}
A stronger notion than validity is that of frame correspondence. The following definition mimics the notion of frame correspondence in modal logic.
\begin{definition}
\label{DEF:AGMcorrespondence}
We say that an axiom $A$ of belief change functions \emph{ is characterized by}, or \emph{corresponds to}, or \emph{characterizes}, a property $P$ of frames if the following is true:
\begin{enumerate}[label=(\arabic*)]
\item axiom $A$ is valid on the class of frames that satisfy property $P$, and
\item if a frame does not satisfy property $P$ then axiom $A$ is not valid on that frame, that is, there is a model based on that frame and a state in that model where the partial belief change function defined by \eqref{RI} violates axiom $A$.
    \end{enumerate}
\end{definition}
The table of Figure \ref{Fig_2} shows, for every AGM axiom, the characterizing property of frames. In \cite{Bon25a} it is shown that the class $\mathcal F_{\mathlarger{\ast}}$ of frames that satisfy the properties listed in Figure \ref{Fig_2} characterizes the set of AGM  belief revision functions  in the following sense (for simplicity we omit the subscript $M$ that refers to a given model; e.g. we write $\Vert\phi\Vert$ instead of $\Vert\phi\Vert_M$):
\begin{enumerate}
\item[(A)] For  every model based on a frame in $\mathcal F_{\mathlarger{\ast}}$ and for every state $s$ in that model, the belief change function $\circ$ (based on $K_s=\{\phi\in\Phi_0:\mathcal B(s)\subseteq \Vert\phi\Vert\}$ and $\Psi=\{\phi\in\Phi_0:\Vert\phi\Vert\ne\varnothing\}$) defined by \eqref{RI} can be extended to a full-domain belief revision function $\ast$ that satisfies the AGM axioms $(K\ast1)$-$(K\ast8)$.
\item[(B)] Let $K\subset \Phi_0$ be a consistent and deductively closed set and let $\ast:\Phi_0\to 2^{\Phi_0}$ be a belief revision function based on $K$ that satisfies the AGM axioms $(K\ast1)$-$(K\ast8)$. Then there exists a frame in $\mathcal F_{\mathlarger{\ast}}$, a model based on that frame and a state $s$ in that model such that (1) $K=K_s=\{\phi\in\Phi_0:\mathcal B(s)\subseteq \Vert\phi\Vert\}$ and (2) the partial belief change function $\circ$ (based on $K_s$ and and $\Psi=\{\phi\in\Phi_0:\Vert\phi\Vert\ne\varnothing\}$) defined by \eqref{RI}  is such that $K_s\circ\phi=K_s\ast\phi$ for every consistent formula $\phi$.
\end{enumerate}
\noindent
The characterization for $(K\ast 4), (K\ast 7)$ and $(K\ast 8)$ is proved in Propositions 9, 3 and 10, respectively, in \cite{Bon25a}. For the remaining axioms we note the following.
\begin{itemize}
\item Validity of $(K\ast 1)$ on all frames is a consequence of Part (B) of Lemma 1 in  \cite{Bon25a}.
    \item Sufficiency for $(K\ast 2)$ is proved in Item 2 of Proposition 1 in \cite{Bon25a}. For necessity, consider a frame that violates Property $(P\ast 2)$, that is, there exist $s\in S$, $\varnothing\ne E\subseteq S$ and $s'\in \mathcal B(s)$  such that $f(s',E)\not\subseteq E$. Construct a model where, for some atomic sentence $p$, $\Vert p\Vert=E$. Then, since $f(s',\Vert p\Vert)\not\subseteq\Vert p\Vert$ and $s'\in \mathcal B(s)$, $p\notin K_s\ast p$, yielding a violation of $(K\ast 2)$.
    \item The proof for $(K\ast 3)$ is as follows.\\
    (A) Sufficiency. Consider a model based on a frame that satisfies Property $(P\ast 3)$. Let $\phi\in\Phi_0$ be such that $\Vert\phi\Vert\ne\varnothing$ and let $s\in S$ and $\psi\in\Phi_0$ be such that $\psi\in K_s\ast\phi$, that is, for every $s'\in\mathcal B(s)$, $f(s',\Vert\phi\Vert)\subseteq\Vert\psi\Vert$, so that $\bigcup\limits_{x\in\mathcal B(s)} f(x,\Vert\phi\Vert)\subseteq\Vert\psi\Vert$.  Fix an arbitrary $s'\in \mathcal B(s)$. If $s'\notin\Vert\phi\Vert$ (that is, $s'\models\neg\phi$) then $s'\models(\phi\rightarrow\psi)$. If $s'\in\Vert\phi\Vert$, then by Property $(P\ast 3)$, $s'\in \bigcup\limits_{x\in\mathcal B(s)} f(x,\Vert\phi\Vert)$ and hence $s'\in\Vert\psi\Vert$; thus $s'\models(\phi\rightarrow\psi)$. It follows that $(\phi\rightarrow\psi)\in K_s$ and thus $\psi\in K_s+\phi$ (since, by Lemma 2 in \cite{Bon25a}, $K_s$ is deductively closed) .\\[3pt]
        (B) Necessity. Consider a frame that violates $(P*3)$, that is, there exist $s\in S$, $s'\in \mathcal B(s)$ and $E\subseteq S$ such that $s'\in E$ and $s'\notin \bigcup\limits_{x\in\mathcal B(s)} f(x,E)$. Construct a model based on this frame where, for some atomic sentences $p$ and $q$, $\Vert p\Vert=E$ and $\Vert q\Vert=\bigcup\limits_{x\in\mathcal B(s)} f(x,E)=\bigcup\limits_{x\in\mathcal B(s)} f(x,\Vert p\Vert)$.  Then, for every $x\in \mathcal B(s)$, $f(x,\Vert p\Vert)\subseteq\Vert q\Vert$ so that $q\in K_s\ast p$. Since $s'\in\Vert p\Vert$ and $s'\notin \Vert q \Vert$, $s'\not\models (p\rightarrow q)$ from which it follows (since $s'\in \mathcal B(s)$) that $(p\rightarrow q)\notin K_s$ and thus $q\notin K_s+p$. Hence $K_s\ast p\not\subseteq K_s+p$.
        \pagebreak
 \begin{figure}[H]
\begin{center}
$\renewcommand{\arraystretch}{2}\arraycolsep=4pt\begin{array}{*{20}{l}}
\qquad\text{\large AGM axiom}&{}&\qquad\text{\large Frame property}\\
\hline
(K*1)\,\,\,K*\phi=Cn(K*\phi)&\vline&\text{No additional property}\\
\hline
%%%%
(K*2)\,\,\,\phi\in K*\phi &\vline&(P*2)\,\,\begin{array}{l}
\forall s\in S, \forall E\in 2^S\setminus\{\varnothing\}, \forall s'\in \mathcal B(s),\\[-10pt]
 f(s',E)\subseteq E\end{array}\\
\hline
%%%%
(K*3)\,\,\,K*\phi\subseteq K+\phi&\vline&(P*3)\,\,\begin{array}{l}
\forall s\in S, \forall E\in 2^S\setminus\{\varnothing\}, \forall s'\in \mathcal B(s),\\[-10pt]
\text{if }s'\in E \text{ then } s'\in \bigcup\limits_{x\in\mathcal B(s)} f(x,E)\end{array}\\
\hline
%%%%
(K*4)\,\,\,\text{if }\neg\phi\notin K*\phi\text{ then }K\subseteq K*\phi&\vline&(P*4)\,\,\begin{array}{l}\forall s\in S,\forall E\in 2^S\setminus\{\varnothing\}, \\[-10pt]
\text{ if } \mathcal B(s)\cap E\ne\varnothing\text{ then,}\\[-10pt]
\forall s'\in\mathcal B(s), f(s',E)\subseteq\mathcal B(s)\cap E\end{array}\\
\hline
%%%%
(K*5b)\,\,\begin{array}{l}\text{If }\neg\phi\text{ is not a tautology}\\[-10pt]
\text{then }K*\phi\ne\Phi_0
\end{array}&\vline&(P*5)\,\,\begin{array}{l}\forall s\in S,\forall E\in 2^S\setminus \{\varnothing\},\exists s'\in\mathcal B(s) \\[-10pt]\text{such that }f(s',E)\ne\varnothing\end{array}\\
\hline
%%%%
(K*6)\,\,\begin{array}{l}\text{if }\phi\leftrightarrow\psi \text{ is a tautology}\\[-10pt]
\text{then }K*\phi=K*\psi\end{array}&\vline&\text{No additional property}\\
\hline
%%%%
(K*7)\,\,\,K*(\phi\wedge\psi)\subseteq (K*\phi)+\psi&\vline&(P*7)\,\,\begin{array}{l}
\forall s\in S,\forall E,F,G\in 2^S\text{ with }E\cap F\ne\varnothing,\\[-10pt]\text{if, } \forall s'\in \mathcal B(s), \,f(s',E\cap F)\subseteq G,\\[-10pt]
\text{then, } \forall s'\in \mathcal B(s),  \,f(s',E)\cap F\subseteq G
\end{array}\\
\hline
%%%%
(K*8)\,\,\begin{array}{l}\text{If }\neg\psi\notin K*\phi,\text{ then}\\[-10pt]
(K*\phi)+\psi\subseteq K*(\phi\wedge\psi)\end{array}&\vline&(P*8)\,\,\begin{array}{l}
\forall s\in S,\forall E,F\in 2^S\setminus\{\varnothing\}\\[-10pt]
 \text{if }\, \exists \hat s\in\mathcal B(s) \text{ such that }\\[-10pt]f(\hat s,E)\cap F\ne\varnothing,\,
  \text{ then,  } \forall s'\in\mathcal B(s),\\[-10pt]
  f(s',E\cap F)\,\,{\mathlarger{\mathlarger \subseteq}} \bigcup\limits_{x\in\mathcal B(s)} \left(f(x,E)\cap F\right).
 \end{array}
\end{array}$
\end{center}
\caption{Semantic characterization of the AGM axioms.}
\label{Fig_1}
\end{figure}
\pagebreak
        \item For axiom $(K\ast 5b)$ the proof  is as follows.\\
        Sufficiency. Consider a model based on a frame that satisfies Property $(P\ast 5)$. Let $\phi\in\Phi_0$ be such that $\Vert\phi\Vert\ne\varnothing$ (thus $\phi$ is consistent) and fix an arbitrary $s\in S$.  By $(P\ast 5)$ there exists an $s'\in\mathcal B(s)$ such that $f(s',\Vert\phi\Vert)\ne\varnothing$. Let $p\in \texttt{At}$ be an atomic sentence . Than $f(s',\Vert\phi\Vert)\not\subseteq\Vert p\wedge\neg p\Vert=\varnothing$. Hence, since $s'\in\mathcal B(s)$, $(p\wedge\neg p)\notin K_s\ast\phi$ so that $K_s\ast\phi\ne\Phi_0$.\\[3pt]
        Necessity, consider a frame that violates $(P*5)$, that is, there exist $s\in S$ and $\varnothing\ne E\subseteq S$ such that, $\forall s'\in \mathcal B(s)$, $f(s',E)=\varnothing$. Construct a model based on this frame where, for some atomic sentence $p$, $\Vert p\Vert=E\ne\varnothing$. Fix an arbitrary $\phi\in\Phi_0$. Then, $\forall s'\in \mathcal B(s)$, $f(s',\Vert p\Vert)=\varnothing\subseteq\Vert\phi\Vert$, so that, by $(RI)$, $\phi\in K_s\ast p$. Thus $K_s\ast p=\Phi_0$.
    \item Validity of $(K\ast 6)$ on all frames is a consequence of the fact that if $\phi\leftrightarrow\psi$ is a tautology, then, in every model, $\Vert\phi\Vert=\Vert\psi\Vert$.
        \end{itemize}
\section{A modal logic for belief revision}
\label{sec3}
We now introduce a simple language with three modal operators: a unimodal belief operator $B$, a bimodal conditional operator $>$ and the unimodal global operator $\square$. The interpretation of $B\phi$ is "the agent believes $\phi$",  the interpretation of $\phi>\psi$ is "if $\phi$ is (or were) the case then $\psi$ is (or would be) the case" and the interpretation of $\square\phi$ is "$\phi$ is necessarily true".\\
The set $\Phi$ of formulas in the language is defined as follows.
\begin{itemize}
\item Let $\Phi_0$ be the set of Boolean formulas defined in Section \ref{Sec2}.
\item Let $\Phi_>$ be the set of formulas of the form $\phi>\psi$ with $\phi,\psi\in\Phi_0$.\footnote{Thus, for example, $\phi>(\psi>\chi)$ is \emph{not} a formula in $\Phi_>$.}
\item Let $\Phi_1$ be the set of Boolean combinations of formulas in $\Phi_0\cup\Phi_>$.
\item Let $\Phi_B$ be the set of formulas of the form $B\phi$ with $\phi\in \Phi_1$.\footnote{Thus, for example, $B(B\phi\rightarrow\phi)$ and $B\square\phi$ are \emph{not} formulas in $\Phi_B$.}
\item Let $\Phi_\square$ be the set of formulas of the form $\square\phi$ with $\phi\in \Phi_0$.\footnote{Thus, for example, $\square(\phi>\psi)$ is \emph{not} a formula in $\Phi_\square$.}
\item Finally, let $\Phi$ be the set of Boolean combinations of formulas in $\Phi_1\cup\Phi_B\cup\Phi_\square$.
\end{itemize}
\subsection{Kripke-Lewis semantics}
As semantics for this modal logic we take the Kripke-Lewis frames of Definition \ref{DEF:frame}. A model based on a frame is obtained, as before, by adding a valuation $V:\texttt{At}\to 2^S$. The following definition expands Definition \ref{Truth0} by adding validation rules for formulas of the form $\square\phi$, $\phi>\psi$ and $B\phi$.
\begin{definition}
\label{TruthModal}
Truth of a formula $\phi$ at state $s$ in model $M$ (denoted by $s\models_M\phi$) is defined as follows:
\begin{enumerate}
\item if $p\in\texttt{At}$ then $s\models_M p$ if and only if $s\in V(p)$.
\item For $\phi\in\Phi$, $s\models_M\neg\phi$ if and only if $s\not\models_M\phi$.
\item $\phi\in\Phi$, $s\models_M(\phi\vee\psi)$ if and only if $s\models_M\phi$ or $s\models_M\psi$ (or both).
\item For $\phi\in\Phi_0$, $s\models_M\square\phi$ if and only if, $\forall s'\in S$, $s'\models_M\phi$ (thus $s\models_M\neg\square\neg\phi$ if and only if, for some $s'\in S$, $s'\models_M\phi$).
\item For $\phi,\psi\in\Phi_0$, $s\models_M(\phi>\psi)$ if and only if, either
\begin{enumerate}[label=(\alph*)]
\item $s\models_M\square\neg\phi$ (that is, $\Vert\phi\Vert_M=\varnothing$), or
\item $s\models_M\neg\square\neg\phi$ (that is, $\Vert\phi\Vert_M\ne\varnothing$) and, for every $s'\in f(s,\Vert\phi\Vert_M)$, $s'\models_M\psi$ (that is, $ f(s,\Vert\phi\Vert_M)\subseteq\Vert\psi\Vert_M$).\footnote{Recall that, by definition of frame, $f(s,E)$ is defined only if $E\ne\varnothing$.}
\end{enumerate}
\item For $\phi\in\Phi_1$, $s\models_M B\phi$ if and only if, $\forall s'\in\mathcal B(s)$, $s'\models_M\phi$ (that is, $\mathcal B(s)\subseteq \Vert\phi\Vert_M$).
\end{enumerate}
\end{definition}
The definition of validity is as in the previous section.
\begin{definition}
\label{DEF:validity2}
A formula $\phi\in\Phi$ is \emph{valid on a frame $F$} if, for every model $M$ based on that frame and for every state $s$ in that model, $s\models_M\phi$. A formula $\phi\in\Phi$ is \emph{valid on a set of frames $\mathcal F$} if it is valid on every frame $F\in\mathcal F$.
\end{definition}
\subsection{Frame correspondence}
The definition of frame correspondence is the standard definition in modal logic.
\begin{definition}
\label{DEF:MODALcorrespondence}
A formula $\phi\in\Phi$ \emph{ is characterized by}, or \emph{corresponds to}, or \emph{characterizes}, a property $P$ of frames if the following is true:
\begin{enumerate}[label=(\arabic*)]
\item $\phi$ is valid on the class of frames that satisfy property $P$, and
\item if a frame does not satisfy property $P$ then $\phi$ is not valid on that frame.
    \end{enumerate}
\end{definition}
The table in Figure \ref{Fig_2} shows, for every property of frames considered in Figure \ref{Fig_1}, the modal formula that corresponds to it. The proofs of the characterizations results are given in the Appendix.
\begin{figure}[h!]
\begin{center}
$\renewcommand{\arraystretch}{2}\arraycolsep=4pt\begin{array}{*{20}{l}}
\qquad\text{\large Frame property}&{}&\text{\large Modal formula }\,\,(\text{for }\phi,\psi,\chi\in\Phi_0)\\
\hline
%%%%
(P*2)\,\,\begin{array}{l}
\forall s\in S, \forall E\in 2^S\setminus\{\varnothing\}, \forall s'\in \mathcal B(s),\\[-10pt]
 f(s',E)\subseteq E\end{array}&\vline&B(\phi>\phi)\\
\hline
%%%%
(P*3)\,\,\begin{array}{l}
\forall s\in S, \forall E\in 2^S\setminus\{\varnothing\}, \forall s'\in \mathcal B(s),\\[-10pt]
\text{if }s'\in E \text{ then } s'\in \bigcup\limits_{x\in\mathcal B(s)} f(x,E)\end{array}&\vline&\big(\neg\square\neg\phi\wedge B(\phi>\psi\big)\rightarrow B(\phi\rightarrow\psi)\\
\hline
%%%%
(P*4)\,\,\begin{array}{l}\forall s\in S,\forall E\in 2^S\setminus\{\varnothing\}, \\[-10pt]
\text{ if } \mathcal B(s)\cap E\ne\varnothing\text{ then,}\\[-10pt]
\forall s'\in\mathcal B(s), f(s',E)\subseteq\mathcal B(s)\cap E\end{array}&\vline&\big(\neg B\neg\phi\wedge B(\phi\rightarrow\psi)\big)\rightarrow B(\phi>\psi)\\
\hline
%%%%
(P*5)\,\,\begin{array}{l}\forall s\in S,\forall E\in 2^S\setminus \{\varnothing\},\exists s'\in\mathcal B(s),\\[-10pt]\text{such that }f(s',E)\ne\varnothing\end{array}&\vline&\big(\neg\square\neg\phi\wedge B(\phi>\psi)\big)\rightarrow \neg B(\phi>\neg\psi)\\
\hline
%%%%
(P*7)\,\,\begin{array}{l}
\forall s\in S,\forall E,F,G\in 2^S\text{ with }E\cap F\ne\varnothing,\\[-10pt]\text{if, } \forall s'\in \mathcal B(s), \,f(s',E\cap F)\subseteq G,\\[-10pt]
\text{then, } \forall s'\in \mathcal B(s),  \,f(s',E)\cap F\subseteq G
\end{array}&\vline&\begin{array}{l}\big(\neg\square\neg(\phi\wedge\psi)\,\wedge\,B((\phi\wedge\psi)>\chi)\big)\\
\rightarrow B\big(\phi>(\psi\rightarrow\chi)\big)\end{array}\\
\hline
%%%%
(P*8)\,\,\begin{array}{l}
\forall s\in S,\forall E,F\in 2^S\setminus\{\varnothing\}\\[-10pt]
 \text{if }\, \exists \hat s\in\mathcal B(s) \text{ such that }\\[-10pt]f(\hat s,E)\cap F\ne\varnothing  \text{ then,  } \forall s'\in\mathcal B(s),\\[-10pt]
  f(s',E\cap F)\,\,{\mathlarger{\mathlarger \subseteq}} \bigcup\limits_{x\in\mathcal B(s)} \left(f(x,E)\cap F\right).
 \end{array}&\vline&\begin{array}{l}\neg B(\phi>\neg\psi)\wedge B(\phi>(\psi\rightarrow\chi))\\[-10pt]\rightarrow\, B\big((\phi\wedge\psi)>(\psi\wedge\chi)\big)\end{array}
\end{array}$
\end{center}
\caption{Modal characterization of the frame properties of Figure \ref{Fig_1}.}
\label{Fig_2}
\end{figure}

\par
Putting together Figures \ref{Fig_1} and \ref{Fig_2}, we have a modal-logic characterization of AGM axioms $(K\ast 2)$, $(K\ast 3)$, $(K\ast 4)$, $(K\ast 5b)$, $(K\ast 7)$ and $(K\ast 8)$. For instance, the modal axiom $B(\phi>\phi)$ corresponds to AGM axiom $(K\ast 2)$ ($\phi\in K\ast\phi$) in the following sense: $(K\ast 2)$ is characterized by frame Property $(P\ast 2)$ which, in turn, characterizes $B(\phi>\phi)$; in other words, $B(\phi>\phi)$ can be viewed as a translation into our modal logic of AGM axiom $(K\ast 2)$.
\par
To complete the analysis we  need to account for AGM axioms $(K\ast 1)$, $(K\ast 5a)$ and $(K\ast 6)$.
\begin{itemize}
\item The modal counterpart of $(K\ast 1)$ ($K\ast\phi=Cn(K\ast\phi)$) can be taken to be the following axiom: for $\phi,\psi,\chi\in\Phi_0$,
    $$\big(B(\phi>\psi)\wedge B(\phi>(\psi\rightarrow\chi))\big)\rightarrow B(\phi>\chi)$$ which is valid on all the frames considered in this paper.\footnote{Fix an arbitrary model, an arbitrary state $s$ and arbitrary $\phi,\psi,\chi\in\Phi_0$ and suppose that $s\models B(\phi>\psi)\wedge B(\phi>(\psi\rightarrow\chi))$. If $\Vert\phi\Vert=\varnothing$ then, by (a) of Item 5 of Definition \ref{TruthModal}, $\forall s'\in\mathcal B(s)$, $s'\models\phi>\chi$ and thus $s\models B(\phi>\chi)$. If $\Vert\phi\Vert\ne\varnothing$, then, $\forall s'\in\mathcal B(s)$, $f(s',\Vert\phi\Vert)\subseteq \Vert\psi\Vert$ [because $s\models B(\phi>\psi)$] and $f(s',\Vert\phi\Vert)\subseteq \Vert\psi\rightarrow\chi\Vert=(S\setminus\Vert\psi\Vert)\cup \Vert\chi\Vert$ [because $s\models B(\phi>(\psi\rightarrow\chi))$]. Since  $\Vert\psi\Vert\,\cap\,\big((S\setminus\Vert\psi\Vert)\cup \Vert\chi\Vert\big)=\Vert\psi\Vert\cap\Vert\chi\Vert\subseteq\Vert\chi\Vert$, we have that, $\forall s'\in\mathcal B(s)$, $f(s',\Vert\phi\Vert)\subseteq \Vert\chi\Vert$, that is, $s\models B(\phi>\chi$).}.
\item AGM axiom $(K\ast 5a)$ (if $\phi$ is a contradiction, then $K\ast\phi=\Phi_0$) can be captured in our logic by the following rule of inference: if $\neg\phi$ is a tautology (a theorem of propositional calculus) then $B(\phi>\psi)$ is a theorem.
\item AGM axiom $(K\ast 6)$ (if $\phi\leftrightarrow\psi$ is a tautology then $K\ast\phi=K\ast\psi$) can be captured in our logic by the following rule of inference: if $\phi\leftrightarrow\psi$ is a tautology then  $B(\phi>\chi)\leftrightarrow B(\psi>\chi)$ is a theorem (for $\phi,\psi,\chi\in\Phi_0$), which is clearly validity preserving in every model, since  if $\phi\leftrightarrow\psi$ is a tautology then $\Vert\phi\Vert=\Vert\psi\Vert$.
\end{itemize}
The table in Figure \ref{Fig_3} synthesizes Figures \ref{Fig_1} and \ref{Fig_2} by showing the correspondence between each AGM axiom and its modal counterpart.
\section{Related literature}
A connection between conditional logic and AGM belief revision has been pointed out in several contributions, in particular \cite{Gioetal98,Gioetal01,Gioetal05,GunSis22}. Although there are some similarities between our approach and those contributions, there are also some important differences. For a detailed discussion see \cite{Bon25a}.
\pagebreak
\begin{figure}[H]
\begin{center}
$\renewcommand{\arraystretch}{2}\arraycolsep=4pt\begin{array}{*{20}{l}}
\qquad\text{\large AGM axiom}&{}&\begin{array}{c}\text{\large Modal axiom / Rule of Inference}\\[-10pt]\,(\text{for }\phi,\psi,\chi\in\Phi_0)\end{array}\\
\hline
(K*1)\,\,\,K*\phi=Cn(K*\phi)&\vline&
\begin{array}{l}B(\phi>\psi)\wedge B(\phi>(\psi\rightarrow\chi))\\[-10pt]
\rightarrow B(\phi>\chi)\end{array}\\
\hline
%%%%
(K*2)\,\,\,\phi\in K*\phi &\vline&
B(\phi>\phi)\\
\hline
%%%%
(K*3)\,\,\,K*\phi\subseteq K+\phi&\vline&
\big(\neg\square\neg\phi\wedge B(\phi>\psi\big)\rightarrow B(\phi\rightarrow\psi)\\
\hline
%%%%
(K*4)\,\,\,\text{if }\neg\phi\notin K*\phi\text{ then }K\subseteq K*\phi&\vline&
\big(\neg B\neg\phi\wedge B(\phi\rightarrow\psi)\big)\rightarrow B(\phi>\psi)\\
\hline
%%%%
(K*5a)\,\,\begin{array}{l}\text{If }\neg\phi\text{ is a tautology, then }\\[-10pt]K*\phi=\Phi_0\end{array}&\vline&
\begin{array}{l} \text{Rule of inference: if  }\neg\phi \text{ is a tautology}\\[-10pt]
\text{then } B(\phi>\psi) \text{ is a theorem} \end{array}\\
(K*5b)\,\,\begin{array}{l}\text{If }\neg\phi\text{ is not a tautology}\\[-10pt]
\text{then }K*\phi\ne\Phi_0\end{array}&\vline&
\big(\neg\square\neg\phi\wedge B(\phi>\psi)\big)\rightarrow \neg B(\phi>\neg\psi)\\
\hline
%%%%
(K*6)\,\,\begin{array}{l}\text{if }\phi\leftrightarrow\psi \text{ is a tautology}\\[-10pt]
\text{then }K*\phi=K*\psi\end{array}&\vline&
\begin{array}{l} \text{Rule of inference: if  }\phi\leftrightarrow\psi \text{ is a tautology}\\[-10pt]
\text{then } B(\phi>\chi)\leftrightarrow B(\psi>\chi) \text{ is a theorem} \end{array}\\
\hline
%%%%
(K*7)\,\,\,K*(\phi\wedge\psi)\subseteq (K*\phi)+\psi&\vline&
\begin{array}{l}\big(\neg\square\neg(\phi\wedge\psi)\,\wedge\,B((\phi\wedge\psi)>\chi)\big)\\[-10pt] \rightarrow B\big(\phi>(\psi\rightarrow\chi)\big)\end{array}\\
\hline
%%%%
(K*8)\,\,\begin{array}{l}\text{If }\neg\psi\notin K*\phi,\text{ then}\\[-10pt]
(K*\phi)+\psi\subseteq K*(\phi\wedge\psi)\end{array}&\vline&
\begin{array}{l}\neg B(\phi>\neg\psi)\wedge B(\phi>(\psi\rightarrow\chi))\\[-10pt]\rightarrow\, B\big((\phi\wedge\psi)>(\psi\wedge\chi)\big)\end{array}
\end{array}$
\end{center}
\caption{The correspondence between AGM axioms and their modal counterparts.}
\label{Fig_3}
\end{figure}
\newpage
\section{Appendix}
\appendix
\begin{proposition}
\label{forK*2}
The modal axiom
\begin{equation}
\label{A2}\tag{A2}
B(\phi>\phi)\qquad(\phi\in\Phi_0)
\end{equation}
is characterized by the following property of frames:
\begin{equation}
\label{P*2}\tag{$P\ast 2$}
\forall s\in S, \forall E\in 2^S\setminus\{\varnothing\}, \forall s'\in \mathcal B(s),\,
 f(s',E)\subseteq E.
\end{equation}
\end{proposition}
\begin{proof}
First we show that Axiom \eqref{A2} is valid on every frame that satisfies  Property \eqref{P*2}. Fix a model based on a frame that satisfies Property \eqref{P*2}, arbitrary $s\in S$, $\phi\in\Phi_0$ and $s'\in\mathcal B(s)$; we need to show that $s'\models(\phi>\phi)$. If $\Vert\phi\Vert=\varnothing$, then, by (a) of Item 5 of Definition \ref{TruthModal}, $s'\models(\phi>\phi )$. If $\Vert\phi\Vert\ne\varnothing$ then, by Property \eqref{P*2}, $f(s',\Vert\phi\Vert)\subseteq\Vert\phi\Vert$ and thus, by (b) of Item 5 of Definition \ref{TruthModal}, $s'\models(\phi>\phi )$.\smallskip
\par
Next we show that Axiom \eqref{A2} is not valid on a frame that violates Property \eqref{P*2}. Fix such a frame, that is, a frame where there exist $s\in S$, $\varnothing\ne E\subseteq S$ and $s'\in\mathcal B(s)$  such that $ f(s',E)\not\subseteq E$. Let $p\in\texttt{At}$ be an atomic formula and construct a model based on this frame where $\Vert p\Vert=E\ne\varnothing$. Then, since $f(s',\Vert p\Vert)\not\subseteq \Vert p\Vert$,  $s'\not\models( p>p)$ and thus $s\not\models B(p>p)$ so that axiom \eqref{A2} is not valid on the given frame.
\end{proof}
\bigskip
%%%%***************************************************************
%%%%%%%%%%%%%%%%%%%%
\begin{proposition}
\label{forK*3}
The modal axiom
\begin{equation}
\label{A3}\tag{A3}
\big(\neg\square\neg\phi\wedge B(\phi>\psi\big)\rightarrow B(\phi\rightarrow\psi)\qquad(\phi,\psi\in\Phi_0)
\end{equation}
is characterized by the following property of frames:
\begin{equation}
\label{P*3}\tag{$P\ast 3$}
\forall s\in S, \forall E\in 2^S\setminus\{\varnothing\}, \forall s'\in \mathcal B(s),\,
\text{if }s'\in E \text{ then } s'\in \bigcup\limits_{x\in\mathcal B(s)} f(x,E).
\end{equation}
\end{proposition}
\begin{proof}
First we show that Axiom \eqref{A3} is valid on every frame that satisfies  Property \eqref{P*3}. Fix a model based on a frame that satisfies Property \eqref{P*3}, arbitrary $s\in S$ and $\phi,\psi\in\Phi_0$ and suppose that $s\models\neg\square\neg\phi\wedge B(\phi>\psi)$. We need to show that  $s\models B(\phi\rightarrow\psi)$. Since $s\models\neg\square\neg\phi$, $\Vert\phi\Vert\ne\varnothing$. Thus, since $s\models B(\phi>\psi)$, $\forall s'\in\mathcal B(s), \,f(s',\Vert\phi\Vert)\subseteq\Vert\psi\Vert$. It follows that
\begin{equation}
\label{3.1}
\bigcup\limits_{x\in\mathcal B(s)} f(x,\Vert\phi\Vert)\subseteq\Vert\psi\Vert.
\end{equation}
Fix an arbitrary $s'\in\mathcal B(s)$. Is $s'\notin \Vert\phi\Vert$ then $s'\models\phi\rightarrow\psi$. If $s'\in \Vert\phi\Vert$, then by Property \eqref{P*3}, $s'\in \bigcup\limits_{x\in\mathcal B(s)} f(x,\Vert\phi\Vert)$ and thus, by \eqref{3.1}, $s'\in\Vert\psi\Vert$, so that $s'\models\phi\rightarrow\psi$. Hence $s\models B(\phi\rightarrow\psi)$.\smallskip
\par
Next we show that Axiom \eqref{A3} is not valid on a frame that violates Property \eqref{P*3}. Fix such a frame, that is, a frame where there exist $s\in S$, $\varnothing\ne E\subseteq S$ and $s'\in\mathcal B(s)$ such that $s'\in E$ but $s'\notin \bigcup\limits_{x\in\mathcal B(s)} f(x,E)$. Let $p,q\in\texttt{At}$ and construct a model based on this frame where $\Vert p\Vert=E$ and $\Vert q\Vert= \bigcup\limits_{x\in\mathcal B(s)} f(x,E)=\bigcup\limits_{x\in\mathcal B(s)} f(x,\Vert p\Vert)$. Then $s'\models p$ but $s'\not\models q$, so that $s'\not\models p\rightarrow q$, from which it follows that $s\not\models B(p\rightarrow q)$, that is, $s\models \neg B(p\rightarrow q)$. To obtain a violation of \eqref{A3} it only remains to show that $s\models\neg\square\neg p\wedge B(p>q)$.  That $s\models\neg\square\neg p$ is a consequence of the fact that, by hypothesis, $\varnothing\ne E=\Vert p\Vert$. Furthermore, since $\Vert p\Vert\ne\varnothing$ and $\Vert q\Vert= \bigcup\limits_{x\in\mathcal B(s)} f(x,\Vert p\Vert)$, for every $y\in \mathcal B(s)$, $f(y,\Vert p\Vert)\subseteq \Vert q\Vert$ and therefore $y\models (p>q)$, so that $s\models B(p>q)$.
\end{proof}
\bigskip
%%%%***************************************************************
%%%%%%%%%%%%%%%%%%%%
\begin{proposition}
\label{forK*4}
The modal axiom
\begin{equation}
\label{A4}\tag{A4}
\big(\neg B\neg\phi\wedge B(\phi\rightarrow\psi)\big)\rightarrow B(\phi>\psi)\qquad(\phi,\psi\in\Phi_0)
\end{equation}
is characterized by the following property of frames:
\begin{equation}
\label{P*4}\tag{$P\ast 4$}
\forall s\in S,\forall E\in 2^S\setminus\{\varnothing\},\,
\text{ if } \mathcal B(s)\cap E\ne\varnothing\text{ then, }
\forall s'\in\mathcal B(s), f(s',E)\subseteq\mathcal B(s)\cap E.
\end{equation}
\end{proposition}
\begin{proof}
First we show that Axiom \eqref{A4} is valid on every frame that satisfies  Property \eqref{P*4}. Fix a model based on a frame that satisfies Property \eqref{P*4}, an arbitrary $s\in S$ and arbitrary $\phi,\psi\in\Phi_0$ and suppose that $s\models\neg B\neg\phi\wedge B(\phi\rightarrow\psi)$. We need to show that $s\models B(\phi>\psi)$. Since $s\models\neg B\neg\phi$, there exists an $s'\in\mathcal B(s)$ such that $s'\models\phi$, that is, $\mathcal B(s)\cap\Vert\phi\Vert\ne\varnothing$. Thus, by Property \eqref{P*4},
\begin{equation}
\label{4.1}
\forall s'\in\mathcal B(s),\, f(s',\Vert\phi\Vert)\subseteq\mathcal B(s)\cap\Vert\phi\Vert.
\end{equation}
Since $s\models B(\phi\rightarrow\psi)$, $\mathcal B(s)\subseteq\Vert\phi\rightarrow\psi\Vert=\Vert\neg\phi\Vert\cup\Vert\psi\Vert$. Hence
\begin{equation}
\label{4.2}
\mathcal B(s)\cap\Vert\phi\Vert\subseteq\big(\Vert\neg\phi\Vert\cup\Vert\psi\Vert\big)\cap\Vert\phi\Vert=\Vert\phi\Vert\cap\Vert\psi\Vert\subseteq\Vert\psi\Vert.
\end{equation}
It follows from \eqref{4.1} and \eqref{4.2} that, $\forall  s'\in\mathcal B(s),\, f(s',\Vert\phi\Vert)\subseteq\Vert\psi\Vert$, that is, $s'\models(\phi>\psi)$ and thus $s\models B(\phi>\psi)$.
\par
Next we show that Axiom \eqref{A4} is not valid on a frame that violates Property \eqref{P*4}. Fix such a frame, that is, a frame where there exist $s\in S$, $\hat s\in \mathcal B(s)$ and $E\in 2^S\setminus\{\varnothing\}$ such that $\mathcal B(s)\cap E\ne\varnothing$ and $f(\hat s,E)\not\subseteq\mathcal B(s)\cap E$. Let $p,q\in\texttt{At}$ and construct a model based on this frame where $\Vert p\Vert=E$ and $\Vert q\Vert=\mathcal B(s)\cap E$. Then $f(\hat s,\Vert p\Vert)\not\subseteq\Vert q\Vert$, that is, $\hat s\not\models p>q$ and thus $s\not\models B(p>q)$, that is,
\begin{equation}
\label{4.3}
s\models\neg B(p>q).
\end{equation}
Since $\mathcal B(s)\cap\Vert p\Vert\ne\varnothing$,
\begin{equation}
\label{4.4}
s\models\neg B\neg p.
\end{equation}
Finally, since $\mathcal B(s)=\big(\mathcal B(s)\cap (S\setminus \Vert p\Vert)\big)\bigcup\big(\mathcal B(s)\cap \Vert p\Vert\big)\subseteq (S\setminus \Vert p\Vert)\cup \big(\mathcal B(s)\cap \Vert p\Vert\big)=\Vert\neg p\Vert\cup\Vert q\Vert=\Vert p\rightarrow q\Vert$,
\begin{equation}
\label{4.5}
s\models B(p\rightarrow q).
\end{equation}
From \eqref{4.3}, \eqref{4.4} and \eqref{4.5} we get a violation of Axiom \eqref{A4}.
\end{proof}
\bigskip
%%%%***************************************************************
%%%%%%%%%%%%%%%%%%%%
\begin{proposition}
\label{forK*5}
The modal axiom
\begin{equation}
\label{A5}\tag{A5}
\big(\neg\square\neg\phi\wedge B(\phi>\psi)\big)\rightarrow \neg B(\phi>\neg\psi)\qquad(\phi,\psi\in\Phi_0)
\end{equation}
is characterized by the following property of frames:
\begin{equation}
\label{P*5}\tag{$P\ast 5$}
\forall s\in S,\forall E\in 2^S\setminus \{\varnothing\},\exists s'\in\mathcal B(s),\text{ such that }f(s',E)\ne\varnothing.
\end{equation}
\end{proposition}
\begin{proof}
First we show that Axiom \eqref{A5} is valid on every frame that satisfies  Property \eqref{P*5}. Fix a model based on a frame that satisfies Property \eqref{P*5}, an arbitrary $s\in S$ and arbitrary $\phi,\psi\in\Phi_0$ and suppose that $s\models\neg \square\neg\phi\wedge B(\phi>\psi)$. Then
\begin{equation}
\label{5.1}
\begin{array}{lll}
(a)&\Vert\phi\Vert\ne\varnothing&(\text{since }s\models \neg \square\neg\phi)\\[3pt]
(b)&\forall s'\in\mathcal B(s), \, s'\models\phi>\psi&(\text{since }s\models B(\phi>\psi)).\end{array}
\end{equation}
Thus, by (a) of \eqref{5.1} and Property \eqref{P*5}, that there exists an $s'\in\mathcal B(s)$, such that $\ f(s',\Vert\phi\Vert)\ne\varnothing$. By (b) of \eqref{5.1}, $f(s',\Vert\phi\Vert)\subseteq\Vert\psi\Vert$, so that (since $f(s',\Vert\phi\Vert)\ne\varnothing$)  $f(s',\Vert\phi\Vert)\not\subseteq\big(S\setminus\Vert\psi\Vert\big)=\Vert\neg \psi\Vert$, that is, $s'\not\models(\phi>\neg\psi)$ and, therefore, $s\not\models B(\phi>\neg\psi)$, that is, $s\models\neg B(\phi>\neg\psi)$.\smallskip
\par
Next we show that Axiom \eqref{A5} is not valid on a frame that violates Property \eqref{P*5}. Fix such a frame, that is, a frame where there exist $s\in S$ and $\varnothing\ne E\subseteq S$ such that, $\forall s'\in\mathcal B(s)$, $f(s',E)=\varnothing$. Construct a model based on this frame where, for some $p,q\in\texttt{At}$, $\Vert p\Vert=E$ and $\Vert q\Vert=\varnothing$. Then (since $\varnothing\ne E=\Vert p\Vert$) $s\models \neg\square\neg p$. Furthermore, $\forall s'\in\mathcal B(s)$, $f(s',\Vert p\Vert)\subseteq \Vert q\Vert$ and $f(s',\Vert p\Vert)\subseteq \Vert\neg q\Vert$. Thus,  $\forall s'\in\mathcal B(s)$,  $s'\models p>q$ and $s'\models p>\neg q$,  so that $s\models B(p>q)$ and  $s\models B(p>\neg q)$ yielding a violation of \eqref{A5}.
\end{proof}
\bigskip
%%%%***************************************************************
%%%%%%%%%%%%%%%%%%%%
\begin{proposition}
\label{forK*7}
The modal axiom
\begin{equation}
\label{A7}\tag{A7}
\big(\neg\square\neg(\phi\wedge\psi)\,\wedge\,B((\phi\wedge\psi)>\chi)\big)\rightarrow B\big(\phi>(\psi\rightarrow\chi)\big)\qquad(\phi,\psi,\chi\in\Phi_0)
\end{equation}
is characterized by the following property of frames:
\begin{equation}
\label{P*7}\tag{$P\ast 7$}
\begin{array}{l}
\forall s\in S,\forall E,F,G\in 2^S\text{ with }E\cap F\ne\varnothing,\\\text{if, } \forall s'\in \mathcal B(s), \,f(s',E\cap F)\subseteq G,
\text{ then, } \forall s'\in \mathcal B(s),  \,f(s',E)\cap F\subseteq G.
\end{array}
\end{equation}
\end{proposition}
\begin{proof}
First we show that Axiom \eqref{A7} is valid on every frame that satisfies  Property \eqref{P*7}. Fix a model based on a frame that satisfies Property \eqref{P*7}, arbitrary $s\in S$ and  $\phi,\psi,\chi\in\Phi_0$ and suppose that $s\models \neg\square\neg(\phi\wedge\psi)\,\wedge\,B((\phi\wedge\psi)>\chi)$. Since $s\models \neg\square\neg(\phi\wedge\psi)$, $\Vert\phi\wedge\psi\Vert\ne\varnothing$ (note that $\Vert\phi\wedge\psi\Vert=\Vert\phi\Vert\cap\Vert\psi\Vert$, so that we also have that $\Vert\phi\Vert\ne\varnothing$ and $\Vert\psi\Vert\ne\varnothing$). Thus, since $s\models B((\phi\wedge\psi)>\chi)$,
\begin{equation}
\label{7.1}
\forall s'\in\mathcal B(s),\, s'\models(\phi\wedge\psi)>\chi,\text{ that is, }f(s',\Vert\phi\Vert\cap\Vert\psi\Vert)\subseteq\Vert\chi\Vert.
\end{equation}
 We need to show that $s\models B\big(\phi>(\psi\rightarrow\chi)\big)$, that is, that, for all $s'\in\mathcal B(s)$, $s'\models\phi>(\psi\rightarrow\chi)$, that is, $f(s',\Vert\phi\Vert)\subseteq\Vert\psi\rightarrow\chi\Vert$. By \eqref{7.1} and Property \eqref{P*7} (with $E=\Vert \phi\Vert$, $F=\Vert\psi\Vert$ and $G=\Vert\chi\Vert$) ,
 \begin{equation}
 \label{7.2}
 \forall s'\in\mathcal B(s), f(s',\Vert\phi\Vert)\,\cap\,\Vert\psi\Vert\subseteq \Vert\chi\Vert\subseteq\big(S\setminus \Vert\psi\Vert\big)\cup\Vert\chi\Vert
 \end{equation}
 which is equivalent to the desired property, since $\big(S\setminus \Vert\psi\Vert\big)\cup\Vert\chi\Vert=\Vert\psi\rightarrow\chi\Vert$.\smallskip
 % \footnote{The following are equivalent: (1) $A\cap B\subseteq C$ and (2) $A\subseteq (S\setminus B)\cup C$. (1) implies (2): pick an arbitrary $x\in A$; if $x\notin B$ then $x\in S\setminus B\subseteq (S\setminus B)\cup C$; if $x\in B$ then by (1) $x\in C\subseteq (S\setminus B)\cup C$. (2) implies (1): intersect both sides of (2) with $B$ to get $A\cap B\subseteq \big(S\setminus B)\cup C\big)\cap B= \big((S\setminus B)\cap B\big)\cup\big(C\cap B\big)=\varnothing\cup\big(C\cap B\big)=C\cap B\subseteq C$. }
 \par
Next we show that Axiom \eqref{A7} is not valid on a frame that violates Property \eqref{P*7}. Fix such a frame, that is, a frame where there exist $s\in S$ and $E,F,G\in 2^S$ with $E\cap F\ne\varnothing$ such that
\begin{equation}
\label{7.3}
\begin{array}{l}
\forall s'\in \mathcal B(s),\,f(s',E\cap F)\subseteq G, \text{ but}\\
\exists \hat s\in  \mathcal B(s) \text{ such that } f(\hat s,E)\cap F\not\subseteq G.
\end{array}
\end{equation}
Let $p,q,r\in\texttt{At}$ and construct a model based on this frame where $\Vert p\Vert=E$, $\Vert q\Vert=F$ and $\Vert r\Vert=G$. Since $E\cap F\ne\varnothing$,
\begin{equation}
\label{7.4}
s\models\neg\square\neg(p\wedge q)
\end{equation}
Furthermore, by \eqref{7.3}, $\forall s'\in \mathcal B(s),\,f(s',\Vert p\Vert\cap\Vert q\Vert)\subseteq \Vert r\Vert$, that is (since $\Vert p\Vert\cap\Vert q\Vert=\Vert p\wedge q\Vert$), $s'\models (p\wedge q)>r$. Hence
\begin{equation}
\label{7.5}
s\models B\big((p\wedge q)>r\big).
\end{equation}
By  \eqref{7.3}, $f(\hat s,\Vert p\Vert)\cap \Vert q\Vert\not\subseteq  \Vert r\Vert$, which is equivalent to $f(\hat s,\Vert p\Vert)\not\subseteq\big(S\setminus\Vert q\Vert)\cup \Vert r\Vert=\Vert q\rightarrow r\Vert$,
%\footnote{Proof that (1) $A\cap B\not\subseteq C$ ir equivalent to (2) $A\not\subseteq \neg B\cup C$. (1) implies (2). From (1) we get that $\exists x\in A\cap B$ such that $x\notin C$. Since $x\in B$, $x\notin \neg B$ and thus, since $x\notin C$, $x\notin\neg B\cup C$. (2) implies (1). From (2) we get that $\exists x\in A$ such that $x\notin \neg B\cup C$, that is, $x\in B$ and $x\notin C$. Thus $\x\in A\cap B$ and $x\notin C$. Hence $A\cap B\not\subseteq C$.}
 so that $\hat s\not\models p>(q\rightarrow r)$, from which it follows (since $\hat s\in \mathcal B(s)$) that
\begin{equation}
\label{7.6}
s\models\neg B\big(p>(q\rightarrow r)\big).
\end{equation}
Thus, by \eqref{7.4}, \eqref{7.5} and \eqref{7.6}, axiom \eqref{A7} is not valid on the given frame.
\end{proof}
\bigskip
%%%%***************************************************************
%%%%%%%%%%%%%%%%%%%%
\begin{proposition}
\label{forK*8}
The modal axiom
\begin{equation}
\label{A8}\tag{A8}
\begin{array}{c}
\big(\neg B(\phi>\neg\psi)\,\wedge\, B(\phi>(\psi\rightarrow\chi)\big)
\rightarrow\, B\big((\phi\wedge\psi)>(\psi\wedge\chi)\big)\\[5pt] \qquad(\phi,\psi,\chi\in\Phi_0)\end{array}
\end{equation}
is characterized by the following property of frames:
\begin{equation}
\label{P*8}\tag{$P\ast 8$}
\begin{array}{l}
\forall s\in S,\forall E,F\in 2^S\setminus\{\varnothing\},\\[4pt]
\text{if }\, \exists \hat s\in\mathcal B(s) \text{ such that }f(\hat s,E)\cap F\ne\varnothing\text{ then,  } \\[4pt]  \forall s'\in\mathcal B(s),
  f(s',E\cap F)\,\,{\mathlarger{\mathlarger \subseteq}} \bigcup\limits_{x\in\mathcal B(s)} \left(f(x,E)\cap F\right).
\end{array}
\end{equation}
\end{proposition}
\begin{proof}
First we show that Axiom \eqref{A8} is valid on every frame that satisfies  Property \eqref{P*8}.
Fix a model based on a frame that satisfies Property \eqref{P*8}, an arbitrary $s\in S$ and arbitrary $\phi,\psi,\chi\in\Phi_0$ and suppose that $s\models\neg B(\phi>\neg\psi)\wedge B(\phi>(\psi\rightarrow\chi))$. We need to show that $s\models B\big((\phi\wedge\psi)>(\psi\wedge\chi)\big)$, that is, that, $\forall s'\in\mathcal B(s), s'\models(\phi\wedge\psi)>(\psi\wedge\chi)$. Since $s\models\neg B(\phi>\neg\psi)$, there exists an $\hat s\in\mathcal B(s)$ such that $\hat s\not\models\phi>\neg\psi$, that is, $\Vert\phi\Vert\ne\varnothing$ and $f(\hat s,\Vert\phi\Vert)\not\subseteq\Vert\neg\psi\Vert$, that is, $f(\hat s,\Vert\phi\Vert)\cap\Vert\psi\Vert\ne\varnothing$, so that, by Property \eqref{P*8} (with $E=\Vert\phi\Vert$ and $F=\Vert\psi\Vert$ and noting that $\Vert\phi\Vert\cap\Vert\psi\Vert=\Vert\phi\wedge\psi\Vert$),
\begin{equation}
\label{8.1}
  \forall s'\in\mathcal B(s),\,f(s',\Vert\phi\wedge\psi\Vert)\,\,{\mathlarger{\subseteq}}\,\, \bigcup\limits_{x \in\mathcal B(s)} {\left(f(x,\Vert\phi\Vert)\cap\Vert\psi\Vert\right)}\,\subseteq\,\Vert\psi\Vert.
\end{equation}
Since $s\models B\big(\phi>(\psi\rightarrow\chi)\big)$, $\forall s'\in\mathcal B(s), f(s',\Vert\phi\Vert)\subseteq\Vert\psi\rightarrow\chi\Vert$. It follows from this and \eqref{8.1} that
\begin{equation}
\label{8.2}
\begin{array}{l}
  \forall s'\in\mathcal B(s),   f(s',\Vert\phi\wedge\psi\Vert)\subseteq \Vert\psi\rightarrow\chi\Vert\cap\Vert\psi\Vert=\Vert\psi\wedge\chi\Vert,\\[5pt]
  \text{that is, }
  s'\models(\phi\wedge\psi)>(\psi\wedge\chi).
  \end{array}
\end{equation}
\par
Next we show that Axiom \eqref{A8} is not valid on a frame that violates Property \eqref{P*8}. Fix such a frame, that is, a frame where there exist $s\in S$, $\hat s,\tilde s\in\mathcal B(s)$ and $E,F\in 2^S\setminus\{\varnothing\}$ such that
\begin{equation}
\label{8.3}
\begin{array}{l}
f(\hat s,E)\cap F\ne\varnothing\,\, \text{ and \,\,} f(\tilde s,E\cap F)\not\subseteq\bigcup\limits_{x \in\mathcal B(s)} {\left(f(x,E) \cap F\right)}.
\end{array}
\end{equation}
Let $p,q,r\in\texttt{At}$ and construct a model based on this frame where $\Vert p\Vert=E$, $\Vert q\Vert = F$ and $\Vert r\Vert=\bigcup\limits_{s' \in\mathcal B(s)} {f(s',E)}$. Then, for all $s'\in\mathcal B(s)$, $f(s',\Vert p\Vert)\subseteq\Vert r\Vert\subseteq\Vert\neg q\Vert\cup\Vert r\Vert=\Vert q\rightarrow r\Vert$, that is, $s'\models p> (q\rightarrow r)$ and thus
\begin{equation}
\label{8.4}
s\models B\big(p>(q\rightarrow r)\big).
\end{equation}
Since, by hypothesis, $f(\hat s,E)\cap F\ne\varnothing$ (that is, $f(\hat s,\Vert p\Vert
)\not\subseteq\Vert\neg q\Vert$, which implies that $\hat s\not\models (p>\neg q$), $s\not\models B(p>\neg q)$, that is,
\begin{equation}
\label{8.5}
s\models\neg B(p>\neg q).
\end{equation}
Furthermore,
 \begin{equation}
\label{8.6}
\bigcup\limits_{s' \in\mathcal B(s)}{\Big(f(s',E)\cap F\Big)}=\left(\bigcup\limits_{s' \in\mathcal B(s)}{f(s',E)}\right)\cap F=\Vert r\Vert\cap\Vert q\Vert=\Vert q\wedge r\Vert.
 \end{equation}
By hypothesis, $f(\tilde s,\Vert p\wedge q\Vert)=f(\tilde s,E\cap F)\not\subseteq\bigcup\limits_{x \in\mathcal B(s)} {\left(f(x,E)\cap F\right)}$ so that, by \eqref{8.6}, $f(\tilde s,\Vert p\wedge q\Vert)\not\subseteq\Vert q\wedge r\Vert$, that is, $\tilde s\not\models(p\wedge q)>(q\wedge r)$ and thus $s\not\models B\big((p\wedge q)>(q\wedge r)\big)$, that is
\begin{equation}
\label{8.7}
s\models\neg B\big((p\wedge q)>(q\wedge r)\big).
\end{equation}
From \eqref{8.4}, \eqref{8.5} and \eqref{8.7} we get a violation of  Axiom \eqref{A8}.
\end{proof}

%%%%%%%%%%%%%
%%%%%%%%%%%%%%%

\end{document}